\documentclass[conference]{IEEEtran}
\IEEEoverridecommandlockouts
\usepackage{cite}
\usepackage{amsmath,amssymb,amsfonts}
\usepackage{algorithmic}
\usepackage{graphicx}
\usepackage{textcomp}
\usepackage{xcolor}
\def\BibTeX{{\rm B\kern-.05em{\sc i\kern-.025em b}\kern-.08em
    T\kern-.1667em\lower.7ex\hbox{E}\kern-.125emX}}

\usepackage{amsmath}
\usepackage{amssymb}
\usepackage{mathtools}
\usepackage{amsthm}
\usepackage{subcaption}
\usepackage{dblfloatfix}
\newcommand{\norm}[1]{\left\lVert#1\right\rVert}

\usepackage{pifont}%
\newcommand{\cmark}{\ding{51}}%
\newcommand{\xmark}{\ding{55}}%
\newcommand{\traj}{\texttt}%

\usepackage{multirow}
\usepackage{booktabs} 

\renewcommand{\paragraph}[1]{\textbf{#1}}

\usepackage[colorlinks=true,allcolors=black,bookmarksdepth=paragraph]{hyperref}

\begin{document}
\bstctlcite{Refs:use_etal} %

\title{wav2pos: Sound Source Localization using \\ Masked Autoencoders \thanks{Corresponding author: Axel Berg (axel.berg@arm.com).
This work was partially supported by the strategic research project \mbox{ELLIIT} and the Wallenberg AI, Autonomous Systems and Software Program (WASP), funded by the Knut and Alice Wallenberg (KAW) Foundation. Model training was enabled by the Berzelius resource provided by the KAW Foundation at the National Supercomputer Centre in Sweden.
 We thank Martin Larsson and Erik Tegler for assistance with data formatting. We also thank Malte Larsson and Gabrielle Flood for their feedback on the paper.}}

\author{\IEEEauthorblockN{Axel Berg$^{* \dagger}$, Jens Gulin$^{* \ddagger}$, Mark O'Connor$^{\S}$, Chuteng Zhou$^{\dagger}$, Karl Åström$^{*}$, Magnus Oskarsson$^{*}$}
\IEEEauthorblockA{\textit{$^{*}$Computer Vision and Machine Learning, Centre for Mathematical Sciences, Lund University}, \textit{$^\dagger$Arm},  \textit{$^{\ddagger}$Sony}, \textit{$^{\S}$Tenstorrent}\\
}
}

\maketitle

\begin{abstract}
We present a novel approach to the 3D sound source localization task for distributed ad-hoc microphone arrays by formulating it as a set-to-set regression problem. By training a multi-modal masked autoencoder model that operates on audio recordings and microphone coordinates, we show that such a formulation allows for accurate localization of the sound source, by reconstructing coordinates masked in the input. Our approach is flexible in the sense that a single model can be used with an arbitrary number of microphones, even when a subset of audio recordings and microphone coordinates are missing. We test our method on simulated and real-world recordings of music and speech in indoor environments, and demonstrate competitive performance compared to both classical and other learning based localization methods.
\end{abstract}

\urlstyle{tt}

\section{Introduction}
\label{introduction}
Mapping, positioning and localization are key enabling technologies for a wide range of applications.
Thanks to its global coverage and scalability, global navigation satellite systems (GNSS) have become
the de-facto standard for outdoor localization. However, for localization in urban areas, indoor
environments and underground, as well as in safety-critical applications, GNSS technology cannot
deliver the accuracy, reliability and coverage needed. Many sensor modalities and setups can be used to address these issues. In this paper we focus our attention on 
sound source localization (SSL), which is the task of determining the location of one or several sound sources using recordings from a microphone array.

Depending on the setup, the sound source position can be estimated in several ways. For fixed equidistant microphones with small physical spacing, localization is typically performed by estimating the direction of arrival (DOA) using e.g.\ steered-response power with phase transform (SRP-PHAT) \cite{diaz2020robust, cho23_interspeech}, spectrograms \cite{wang23j_interspeech, phokhinanan23_interspeech} or raw waveforms \cite{he22_interspeech} as input features.

When the microphone positions are distributed around the sound source in an ad-hoc fashion, it is possible to estimate the 3D location of the sound source with respect to some coordinate system, given the microphone positions. Depending on whether the sound source is time-synchronized with the microphones or not this is known as trilateration or multilateration, respectively.

\begin{figure}[t!]
\centering
\includegraphics[width=\linewidth, trim={0 0 3cm 0},clip]{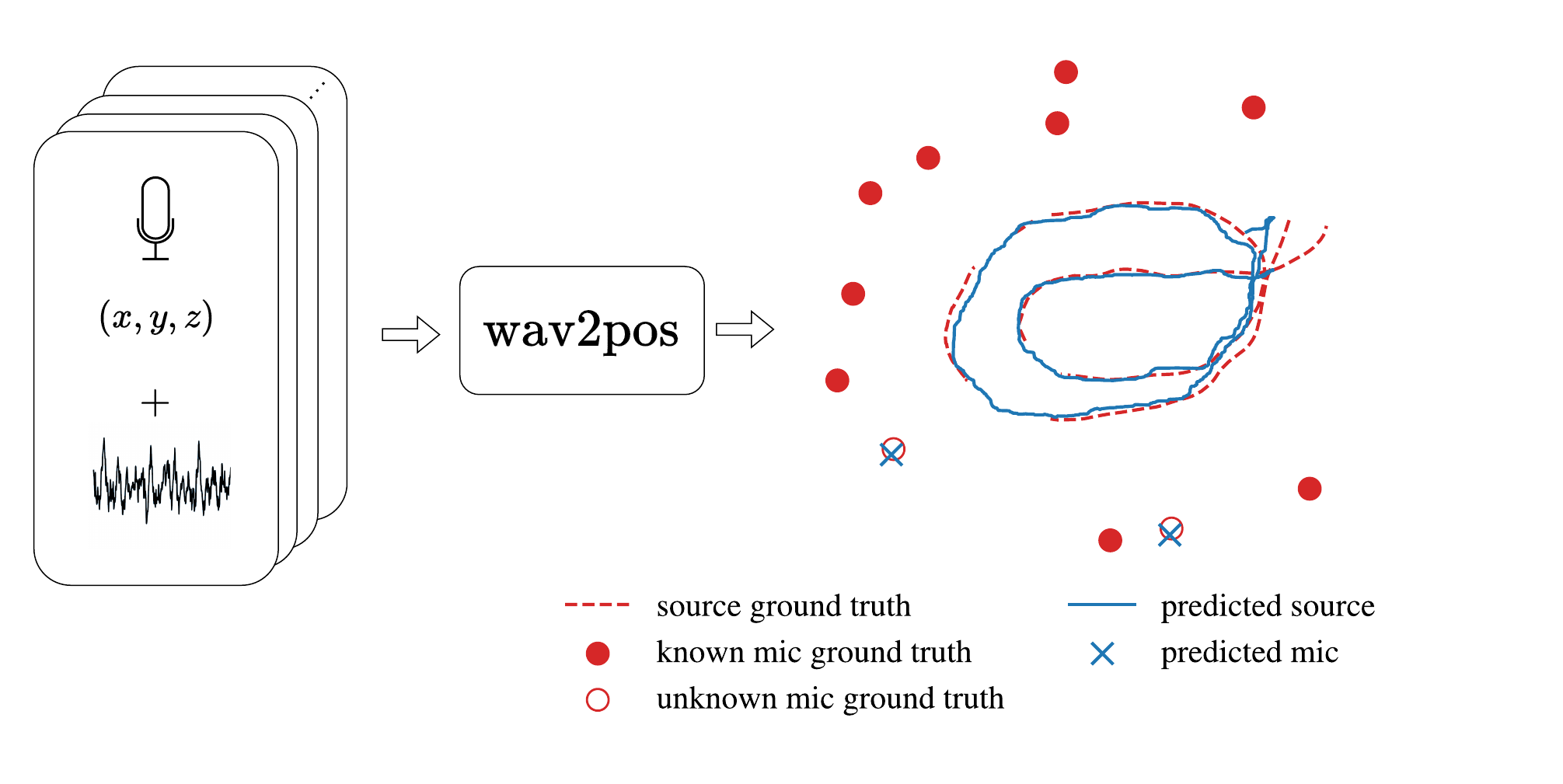}
\caption{Method overview: wav2pos can simultaneously localize a moving sound source and several microphones given audio recordings and microphone coordinates on a frame-by-frame basis. Here, predictions on the \traj{music3} recording from the LuViRa dataset \cite{yaman2023luvira} are shown (viewed from above), where a moving median filter has been applied to predictions for better visualization.}
\label{curve}
\end{figure}

\paragraph{Classical methods.} For trilateration, there is a large body of previous work. The minimal amount of data needed is when the number of distance measurements equals the spatial dimension, and this problem has a closed-form solution \cite{thomas_revisiting_2005, manolakis_efficient_1996, coope_reliable_2000}.
For the over-determined problem, finding the maximum likelihood (ML) estimate given Gaussian noise in the measurements is a nonlinear, non-smooth and non-convex problem.  Unlike the minimal case, there is no closed-form solution. A number of iterative methods, with various convergence guaranties, exist \cite{luke_simple_2017, beck_iterative_2008, jyothi_solvit_2020, sirola_closed-form_2010}.
Simplifications to  the ML problem, include relaxations by minimizing the error in the squared distance measurements  \cite{larsson_optimal_2019, zhou_closed-form_2011, beck_exact_2008, cheung_least_2004}. Various heuristics can be used to arrive at  a linear  formulation, see \cite{stoica2006lecture} and references therein.
For the multilateration problem, classical methods  rely on estimating the time-differences of arrival (TDOA) between pairs of microphones using pairwise feature extractors, where the most common one is the generalized cross-correlation with phase transform (GCC-PHAT) \cite{knapp1976generalized}. TDOA measurements can then be used to perform multilateration, where the sound source location is obtained by solving a system of equations using e.g.\ the least squares method or a minimal solver \cite{aastrom2021extension}.

\paragraph{Learning based methods.} Learning based methods have been used extensively for TDOA  \cite{salvati2021time, berg22_interspeech, raina2023syncnet} and DOA  \cite{diaz2020robust, cho23_interspeech, wang23j_interspeech, phokhinanan23_interspeech, he22_interspeech} estimation. However, there is a lack of research on learning based methods for localization using distributed microphone arrays. Vera-Diaz et al.\ \cite{vera2018towards} used a convolutional neural network to directly regress the source coordinates, but only for a fixed microphone array. Grinstein et al.\ \cite{grinstein2023dual} proposed a dual-input neural network for end-to-end SSL with spatial coordinates in 2D, where both audio signals and microphone coordinates are used as input. This allows the model to be trained in setups with ad-hoc arrays where the microphone locations are not fixed. However, the network is limited to using a fixed number of inputs, which prevents the user from adding or removing microphones to the array at inference time. Similarly, in \cite{grinstein2023graph} a graph neural network is proposed for the same task that works with variable number of microphones by aggregating over GCC-PHAT features, which are used as inputs to the network. Similar architectures have also been proposed in \cite{gong2022end, feng2023soft}, where a transformer \cite{vaswani2017attention} architecture is used for localization on a 2D grid. However, these methods fail to scale to three dimensions since the discretization of large spaces becomes infeasible. For further reading about prior work, we refer the reader to the extensive survey published in \cite{grumiaux2022survey}.

\begin{figure*}[hptb]
\centering
\includegraphics[trim={1cm, 0, 2cm, 0},clip,width=\textwidth]{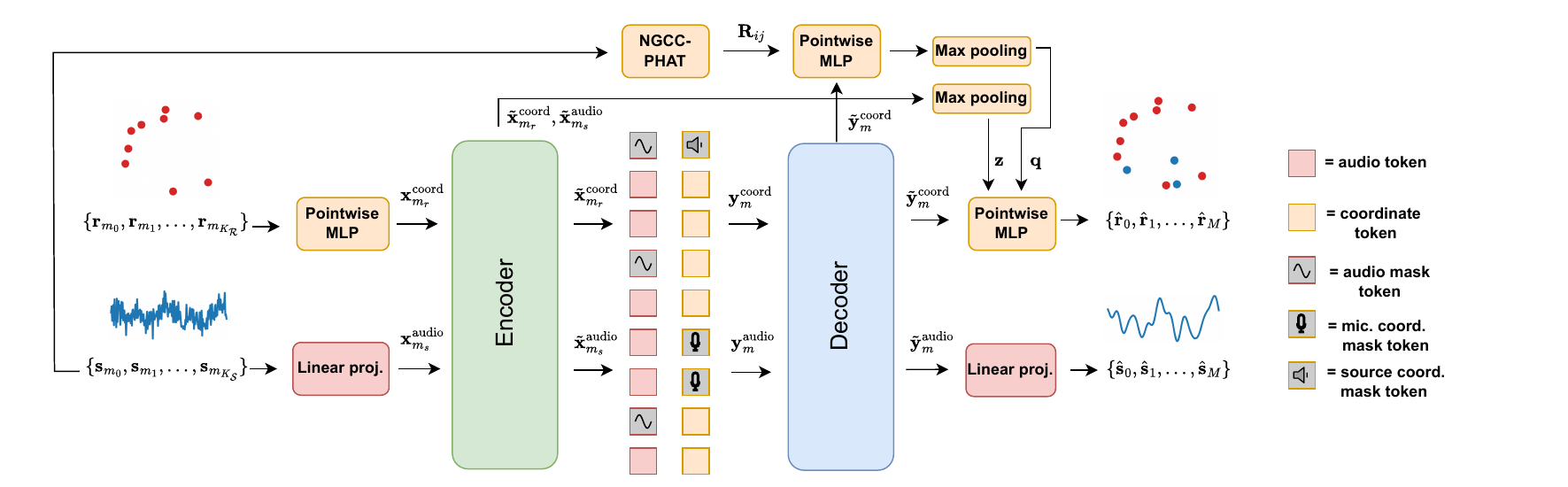}
\caption{High-level illustration of the proposed wav2pos method. Modality embedding (added before encoder and decoder) and pairwise positional encoding (added before decoder) are omitted for brevity. The mask tokens are not generated by the encoder, but appended as learnable tokens in the input sequence to the decoder.}
\label{method-fig}
\end{figure*}

\paragraph{Main contributions.} In this work, we present a novel method for single-source 3D SSL that directly predicts the sound source coordinates using an ad-hoc distributed microphone array. Inspired by the success of masked autoencoders in natural language processing \cite{devlin2018bert}, computer vision \cite{he2022masked}, audio processing \cite{huang2022masked} and combinations thereof \cite{geng2022multimodal}, we formulate the SSL problem as a multi-modal set-to-set regression problem, which allows our method to localize not only the sound source, but also solve a variety of similar problems where audio or locations are missing for some microphones, as shown in Figure \ref{curve}. %

\section{Method}
\paragraph{Problem setup.} Consider $M$ microphones with ad-hoc coordinates $\mathbf{r}_m \in \mathbb{R}^3$, $m = 1, \hdots, M$ and \emph{one} audio source located at $\mathbf{r}_0 \in \mathbb{R}^3$. For a given time slot of $N$ samples, the source emits a signal $\mathbf{s}_0 \in \mathbb{R}^N$ and each microphone receives a delayed and noisy copy $\mathbf{s}_m \in \mathbb{R}^N$, $m = 1, \hdots, M$, of the signal that depends on the impulse response $h_m$ from the source position to each microphone,
\begin{align}
s_m[n] = (h_m * s_0)[n] + w_m[n], \quad n = 1, \hdots, N,
\end{align}  
where $w_m$ can be approximated as i.i.d.\ Gaussian noise and $(*)$ denotes the convolution operator. The window length $N$ is assumed to be small enough for the sound source to be modeled as stationary. The SSL task is to recover the audio source location $\mathbf{r}_0$, given the recordings of each microphone and their known locations. In the more general setup, the task can be extended to predict some unknown microphone locations as well. Note that we only consider a single sound event for each prediction, and finding the trajectory of a moving sound source thus amounts to making predictions on a frame-by-frame basis. In the general, full calibration problem, it is not possible to estimate arbitrary microphone positions from a single sound event using only distance measurements. However, in our simplified problem there is a limited number of possible microphone positions in the training data. Furthermore, other spatial cues such as the acoustic features of the environment can be learnt by the model, which makes the problem setup tractable.

\paragraph{Masked autoencoders for localization.} The main idea of our method is to consider the SSL problem as function approximation over the set $\{\mathbf{s}_m, \mathbf{r}_m\}_{m=0}^M$ of audio signals and locations. This allows for exploiting redundancy in the data by \emph{masking}, where missing inputs are filled in by the model. In this context, masking can be used both as a training strategy that enables it to perform localization using different microphone array setups, and to predict missing microphone coordinates. Note that while masked autoencoders are often trained in a self-supervised manner, our method is fully supervised, but random masking is used during training in order to increase robustness to missing inputs.

Let $\mathcal{S} = \{m: \mathbf{s}_m \text{ not masked} \}$ and $\mathcal{R} = \{m: \mathbf{r}_m \text{ not masked} \}$ denote the set of non-masked recorded audio signals and coordinates respectively, with set sizes $K_\mathcal{S} = |\mathcal{S}|$ and $K_\mathcal{R} = |\mathcal{R}|$. Using the set of non-masked inputs, we seek to learn a function $f_{\boldsymbol{\theta}}$ that outputs predictions $\hat{\mathbf{s}}_m, \hat{\mathbf{r}}_m$ for the complete set:
\begin{align}
\label{set_estimation}
\{\hat{\mathbf{s}}_m, \hat{\mathbf{r}}_m\}_{m=0}^M = f_{\theta} \big( \{\mathbf{s}_{m_s}\}_{m_s \in \mathcal{S}}, \{\mathbf{r}_{m_r}\}_{m_r \in \mathcal{R}} \big).
\end{align}

The function approximation model consists of an encoder, which operates on the non-masked subset of inputs, and a decoder that forms predictions on the entire set. Both the encoder and decoder consist of sequential transformer blocks that process audio and coordinate tokens jointly, as shown in Figure \ref{method-fig}. In other words, both audio and coordinate tokens are treated as elements of the same unordered set by each transformer block.   

We train our method using mean squared error loss on the sound source coordinate, the masked microphone coordinates and the non-masked audio. Reconstructing masked audio patches requires learning impulse responses across the room, which is out of the scope of this work, and initial experiments showed poor performance for that task. For audio prediction, we therefore restrict the loss to the non-masked audio tokens, since this allows the model to learn to perform audio de-noising. Thus, the total loss becomes
\begin{equation}
\begin{aligned}
L =& \frac{\lambda_\text{audio}}{K_\mathcal{S}} \sum_{m \in \mathcal{S}} \norm{\hat{\mathbf{s}}_m - \mathbf{s}_m}^2_2 + \lambda_\text{source}\norm{\hat{\mathbf{r}}_0 - \mathbf{r}_0}^2_2 \\ +& \frac{\lambda_\text{mic}}{M-K_\mathcal{R}} \sum_{m \notin \mathcal{R},  m \geq 1} \norm{\hat{\mathbf{r}}_m - \mathbf{r}_m}^2_2,
\end{aligned}
\end{equation}
where $\lambda_\text{audio}, \lambda_\text{source}, \lambda_\text{mic}$ are hyperparameters that balance the contribution of the source localization error, microphone localization error and audio reconstrucion error, respectively. We will now proceed to describe the method in more detail. 

\paragraph{Feature embedding.} Audio snippets are first processed individually for each microphone using a linear projection in order to form input tokens $\mathbf{x}_{m_s}^{\text{audio}} = \mathbf{W}_{\text{enc}} \mathbf{s}_{m_s}$, where $\mathbf{W}_{\text{enc}}  \in \mathbb{R}^{d \times N}$ and $d$ is the embedding dimension. Similarly, the microphone coordinates are projected to the same embedding space using a point-wise MLP: $\mathbb{R}^3 \rightarrow \mathbb{R}^d$ with one hidden layer and batch normalization \cite{ioffe2015batch} that computes the coordinate tokens $\mathbf{x}_{m_r}^{\text{coord}}$. In order to let the model distinguish between the two modalities of tokens, we follow \cite{geng2022multimodal} and add learnable modality embeddings $\mathbf{v}^{\text{audio}}_\text{enc}, \mathbf{v}^{\text{coord}}_\text{enc} \in \mathbb{R}^d$ to each token according to its modality. The audio and coordinate features are then processed jointly by a series of $D$ sequential transformer encoder blocks with layer normalization \cite{ba2016layer} and GELU activations \cite{hendrycks2016gaussian}. 

After the final encoder layer, two types of learnable mask tokens $\mathbf{u}^{\text{audio}}, \mathbf{u}^{\text{coord}} \in \mathbb{R}^d$ are appended to the output set for each input that was masked out from the input. Additionally a special mask token $\mathbf{u}^{\text{source}} \in \mathbb{R}^d$ for the sound source coordinate is appended, and new modality embeddings for the decoder $\mathbf{v}^{\text{audio}}_\text{dec}, \mathbf{v}^{\text{coord}}_\text{dec} \in \mathbb{R}^d$ are added to all tokens.

\paragraph{Pairwise positional encoding.} Since the source coordinate prediction should be invariant to permutations of the microphone order (and all other outputs should be equivariant), we do not add any form of positional encoding in the usual sense that encodes the relative or absolute ordering of tokens, as is typically done when using transformers for sequence modeling. However, the decoder still needs to be informed about which audio snippet corresponds to which microphone location and vice versa, or whether the corresponding audio/coordinate token was masked. Therefore, we propose a pairwise message-passing embedding that communicates information between tokens originating from the same microphone. The messages are computed using two separate functions $\gamma^{a\rightarrow c}, \gamma^{c\rightarrow a}: \mathbb{R}^d \rightarrow \mathbb{R}^d$ that are implemented as MLPs with a single hidden layer. In total, the inputs $\mathbf{y}_{m}^{\text{audio}}, \mathbf{y}_{m}^{\text{coord}}$ to the decoder are given by
\begin{equation}
\begin{aligned}
\mathbf{y}_{m}^{\text{audio}} &= \mathbf{t}_{m}^{\text{audio}} + \mathbf{v}^{\text{audio}}_\text{dec} + \gamma^{c\rightarrow a}(\mathbf{t}_m^{\text{coord}}),   \\
\mathbf{y}_{m}^{\text{coord}} &= \mathbf{t}_m^{\text{coord}} + \mathbf{v}^{\text{coord}}_\text{dec} + \gamma^{a\rightarrow c}(\mathbf{t}_{m}^{\text{audio}}), 
\end{aligned}
\label{eq:y}
\end{equation}
for $m = 0, \hdots, M$ and where
\begin{align}
\mathbf{t}_m^{\text{audio}} = \begin{cases} \tilde{\mathbf{x}}_{m}^{\text{audio}}, \,  m \in \mathcal{S} \\ \mathbf{u}^{\text{audio}},\, m \notin \mathcal{S} \end{cases}
\mathbf{t}_m^{\text{coord}} = \begin{cases} \tilde{\mathbf{x}}_{m}^{\text{coord}}, \,  m \in \mathcal{R} \\ \mathbf{u}^{\text{source}}, \, m = 0 \\ \mathbf{u}^{\text{coord}}, \, m \notin \mathcal{R} \end{cases},
\end{align}
and $\tilde{\mathbf{x}}_{m}^{\text{audio}}, \tilde{\mathbf{x}}_{m}^{\text{coord}}$ are the encoder outputs. The tokens are then passed through the decoder, which, similarly to the encoder, consists of $D$ sequential transformer layers.
\begin{table*}[t!]
\centering
\caption{Model properties and localization performance on the LuViRA \cite{yaman2023luvira} \traj{music3} and \traj{speech3} trajectories using all 11 microphones. Input types refers to: 1 - GCC-PHAT, 2 - NGCC-PHAT, 3 - raw audio waveforms.}
\begin{tabular}{l|l|cccc|cc|cc}
\toprule
 & 
 & \multicolumn{4}{c}{} & \multicolumn{2}{|c|}{\traj{music3}} & \multicolumn{2}{c}{\traj{speech3}} \\
Setup & Method & Input & var. \#mics & perm.\ inv.\ & \#params & MAE [cm] $\downarrow$ &  acc@30cm $\uparrow$ & MAE [cm] $\downarrow$ & acc@30cm $\uparrow$ \\ \midrule
$1_a$ & Multilat \cite{aastrom2021extension} & 1 & \cmark & \cmark & - & $38.8 \pm 2.5$& $72.5 \pm 1.6$ & $72.8 \pm 4.4$ & $55.7 \pm 2.1$\\
& Multilat* \cite{aastrom2021extension, berg22_interspeech} & 2 & \cmark & \cmark & 0.9M & $16.3 \pm 1.6$ & $\mathbf{94.7 \pm 0.8}$ & $34.9 \pm 3.2$ & $\mathbf{84.9 \pm 1.6}$ \\
& DI-NN \cite{grinstein2023dual} & 3 & \xmark & \xmark & 3.6M & $26.0 \pm 0.8$ & $73.0 \pm 0.2$ & $44.7 \pm 1.7$ & $45.9 \pm 2.3$ \\
& GNN \cite{grinstein2023graph} & 1 & \cmark & \xmark & 2.2M & $ 17.0 \pm 0.7$ & $90.7 \pm 1.0$ & $31.9 \pm 1.6$ & $71.2 \pm 2.0$ \\
& wav2pos & 2+3 & \cmark & \cmark & 10.5M & $\mathbf{14.2 \pm 0.5}$ & $\mathbf{95.4 \pm 0.7}$ & $\mathbf{23.6 \pm 1.2}$ & $81.6 \pm 1.7$ \\ \bottomrule
\end{tabular}
\label{properties}
\end{table*}

At the output of the decoder, features $\tilde{\mathbf{y}}_{m}^{\text{audio}}, \tilde{\mathbf{y}}_{m}^{\text{coord}}$ are collected for all audio sequences and coordinates. Audio reconstructions are formed by using a simple linear projection as $\hat{\mathbf{s}}_m = \mathbf{W}_{\text{dec}}\tilde{\mathbf{y}}_{m}^{\text{audio}}$ for $m \in \mathcal{S}$, where $\mathbf{W}_{\text{dec}} \in \mathbb{R}^{N \times d}$.

\paragraph{Time delay feature module.} For the coordinate predictions, we use additional information from previous layers by computing a global feature $\mathbf{z} \in \mathbb{R}^d$ by max-pooling over all encoder outputs as $\mathbf{z}=\max_{m_s \in \mathcal{S}, m_r \in \mathcal{R}} (\tilde{\mathbf{x}}_{m_s}^{\text{audio}}, \tilde{\mathbf{x}}_{m_r}^{\text{coord}})$. Similarly to the method proposed in \cite{grinstein2023graph}, we also use TDOA features $\mathbf{R}_{ij} \in \mathbb{R}^{2\tau+1}$, which we obtain from a pre-trained NGCC-PHAT \cite{berg22_interspeech} for all non-masked pairs of audio inputs, where $\tau$ is determined by the maximum possible delay between microphones. We then combine each time-delay feature with coordinate features from the two corresponding microphones, and pool over all $M(M-1)$ microphone pairs in order to form a global feature which contains information about all TDOA measurements as
\begin{align}
\mathbf{q} = \max_{\substack{i, j \in \mathcal{S}, i \neq j}} \varphi \left(\mathbf{R}_{ij}, \tilde{\mathbf{y}}_{i}^{\text{coord}}, \tilde{\mathbf{y}}_{j}^{\text{coord}}  \right),
\end{align}
where $\varphi: \mathbb{R}^{2(\tau + d) + 1} \rightarrow \mathbb{R}^{d} $ is a single hidden-layer MLP. In order to form the final predictions, we use two separate MLPs $\psi_\text{source}, \psi_\text{mic}:\mathbb{R}^{3d} \rightarrow \mathbb{R}^3$ that produce coordinate predictions for the sound source and microphones as
\begin{equation}
\begin{aligned}
&\hat{\mathbf{r}}_0 = \psi_\text{source}(\tilde{\mathbf{y}}_{0}^{\text{coord}}, \mathbf{z}, \mathbf{q}), \\ &\hat{\mathbf{r}}_m = \psi_\text{mic}(\tilde{\mathbf{y}}_{m}^{\text{coord}}, \mathbf{z}, \mathbf{q}), \quad m = 1, \hdots, M.
\end{aligned}
\end{equation}

\paragraph{Masking strategy.} When training the model, different masking strategies can be used depending on the use case. If the number of microphones is known to be fixed at inference time, we only mask out the audio and coordinates of the sound source, i.e.\ $\mathcal{R} = \mathcal{S} = \{1, ..., M\}$. When the number of microphones available is not fixed, we instead randomly mask a subset of both the audio snippets and coordinates in order to make the model more robust to using a variable number of microphones. A unique solution to the multilateration problem requires four TDOA measurements from five microphones \cite{gillette2008linear}, and therefore we always restrict masking such that $|\mathcal{S} \cap \mathcal{R}| \geq 5$. We also require that not both audio and coordinates from the same microphone are masked.

\section{Experimental Results}

\paragraph{Real indoor recordings.} We evaluate our method on the LuViRA \cite{yaman2023luvira} audio-only dataset, which contains eight real-world recordings, about one minute long, of music or speech in an indoor environment. Each recording is captured by 11 stationary and synchronized microphones and the speaker location ground truth is given by a motion capture system. An additional 12:th microphone is placed next to the speaker, and can be used as stand-in for the source audio $\mathbf{s}_0$. We evaluate on the \traj{music3} and \traj{speech3} recordings, and use the remaining three music and three speech recordings for training. In order to improve model generalization to unseen source locations, the dataset is also expanded with simulated recordings, where the sound source is randomly sampled in a room of size $7 \times 8 \times 2$ m, with microphones placed in the same positions as in the real recordings. Simulations are done using Pyroomacoustics \cite{scheibler2018pyroomacoustics}, where in each time frame, we randomly sample a source position and a reverberation time $t_{60}$ in the range (0, 0.4) and use audio captured from the 12:th microphone as input to the simulation. The total amount of training data is therefore approximately $2\times (3 + 3) = 12$ minutes of audio recordings.

We initialize all network layers, as well as mask tokens and modality embeddings, from a Gaussian distribution $\mathcal{N}(0, 0.02)$, then train for 500 epochs using the AdamW optimizer \cite{loshchilov2017decoupled} with a batch size of 256, a learning rate of 0.0005 and weight decay of 0.1. In all experiments we use $\lambda_\text{source} = \lambda_\text{mic} = 1.0$ and $\lambda_\text{audio}=0.1$, an embedding dimension $d=256$, $D=4$ transformer layers and a signal length of $N = 2048$ at a sample rate of 16 kHz. For TDOA features, we use NGCC-PHAT \cite{berg22_interspeech} by pre-training it on the same dataset. For data augmentation we use additive Gaussian noise and random time shifts of the audio, uniformly sampled in [-0.05, 0.05] s. The same time shift is applied to all microphones in order to preserve relative time differences, and the speaker is assumed to be stationary within this time period. Silent periods are excluded (for both training and inference) by thresholding the signal power. The audio reconstruction loss is computed on the non-masked inputs without noise, which enables the model to perform de-noising.

We compare our method to a robust multilateration method \cite{aastrom2021extension}, where TDOAs are estimated using GCC-PHAT or a pre-trained NGCC-PHAT. We also extend the dual input neural network (DI-NN) \cite{grinstein2023dual} and graph neural network (GNN) \cite{grinstein2023graph} methods to 3D localization, and train them on the same dataset using the MSE loss, but with the hyperparameters proposed in the corresponding publications. Localization errors are truncated at 3 m, since the traditional multilateration method sometimes yields very large errors or fails to converge.

The results are shown in Table \ref{properties}, along with the input type used by each method, whether they support a variable number of microphones, if they are invariant to permutations of the microphone order and the number of learnable parameters. Evaluation is done assuming all microphone locations are known, which we denote Setup $1_a$. The mean absolute error (MAE) and accuracy are evaluated using a 95 \% confidence interval by bootstrapping. The results indicate that our method consistently has the lowest MAE for both music and speech recordings. Although multilateration with NGCC-PHAT achieves similar accuracy at the 30 cm threshold, our method has a shorter tail in the error distribution and therefore yields a lower MAE.

\paragraph{Evaluation with missing inputs.} In order to test our method in different problem setups, we also train it using random masking. During training, we randomly leave between 8 and 11 coordinates and audio snippets, such that at least 5 microphones are in both sets. Since the sound source audio $\mathbf{s}_0$ might be known in some scenarios, we mask this token with 50 \% probability, and denote our method trained with random masking as wav2pos$_\mathcal{M}$. At inference time, unknown microphone locations are masked and, if the corresponding sound recordings are not masked, the coordinates can be predicted by the decoder.
\begin{table}[t]
\centering
\footnotesize
\caption{Sound source localization MAE [cm] on the \traj{speech3} trajectory using different number of microphones and setups. Multilat* fails to converge for 5 microphones.}
\begin{tabular}{l|l|ccc}
\toprule
Setup & Method &$M=5$ & $M=7$ & $M=9$ \\ \midrule
$1_a$ &Multilat & $244.9 \pm 4.8$ & $133.1 \pm 5.7$ & $94.3 \pm 5.0$ \\
&Multilat* & N/A & $105.6 \pm 6.1$ & $56.7 \pm 4.5$ \\
&DI-NN$_M$ &$94.9 \pm 2.6$ & $76.1 \pm 2.0$ & $58.5 \pm 1.6$  \\
&GNN$_\mathcal{M}$ & $80.5 \pm 2.2$ & $53.1 \pm 1.9$ & $41.1 \pm 1.7$ \\
&wav2pos$_\mathcal{M}$ & $\mathbf{66.8 \pm 2.0}$ & $\mathbf{38.8 \pm 1.7}$ & $\mathbf{28.4 \pm 1.4}$ \\ \midrule
$1_b$ & wav2pos$_\mathcal{M}$ & $42.3 \pm 1.4$ & $26.1 \pm 1.0$ & $20.4 \pm 1.0$ \\ \midrule
$2_a$ & wav2pos$_\mathcal{M}$ & $47.2 \pm 1.8$ & $32.2 \pm 1.4$ & $25.3 \pm 1.2$ \\ \midrule
$2_b$ & wav2pos$_\mathcal{M}$ & $33.0 \pm 1.2 $ & $23.3 \pm 1.0$ & $19.6 \pm 1.0$ \\ \bottomrule
\end{tabular}
\label{mic_ablation}
\end{table}
\begin{table}[t]
\centering
\footnotesize
\caption{Microphone localization MAE [cm] over all unknown microphone locations, on the \traj{speech3} trajectory using different numbers of known microphone locations.}
\begin{tabular}{l|l|ccc} \toprule
Setup & Method &$M=7$ & $M=8$ & $M=9$ \\ \midrule
$2_a$ & wav2pos$_\mathcal{M}$ & $182.8 \pm 1.7$ & $93.1 \pm 1.9$ & $36.8 \pm 1.9$ \\ \midrule
$2_b$ & wav2pos$_\mathcal{M}$ & $181.7 \pm 1.7$ & $90.4 \pm 1.8$ & $34.8 \pm 1.6$ \\ \bottomrule
\end{tabular}
\label{mic_loc}
\end{table}
\begin{figure*}[b]
\centering
        \begin{subfigure}[b]{0.195\linewidth}
                \centering
                \includegraphics[width=\linewidth, trim={0.5cm 0.9cm 0 0.8cm},clip]{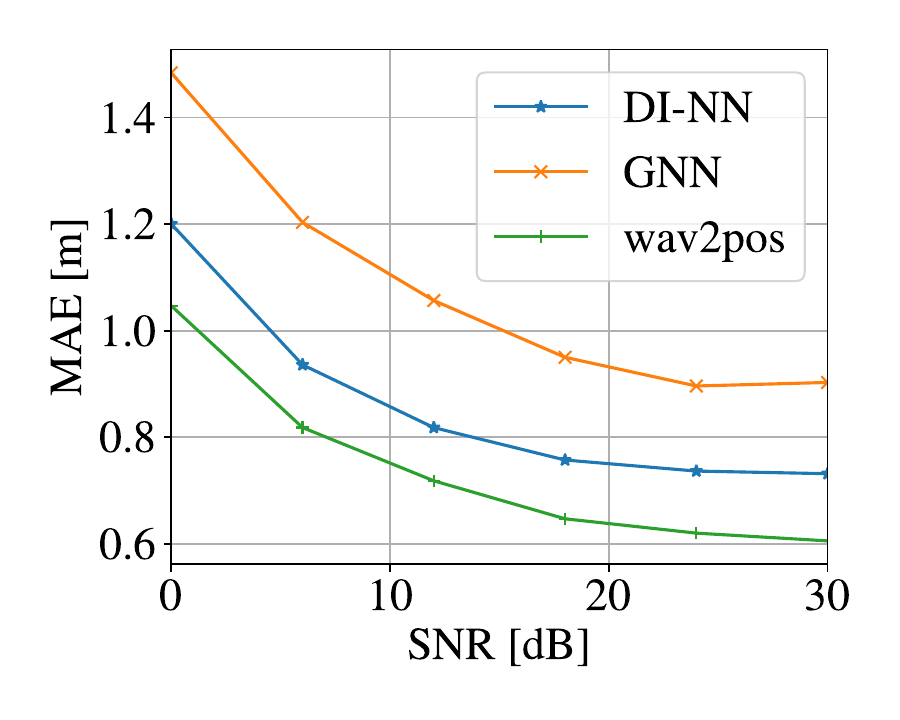}
                \caption{MAE vs SNR}
        \end{subfigure}\hfill
        \begin{subfigure}[b]{0.195\linewidth}
                \centering
                \includegraphics[width=\linewidth, trim={0.5cm 0.9cm 0 0.8cm},clip]{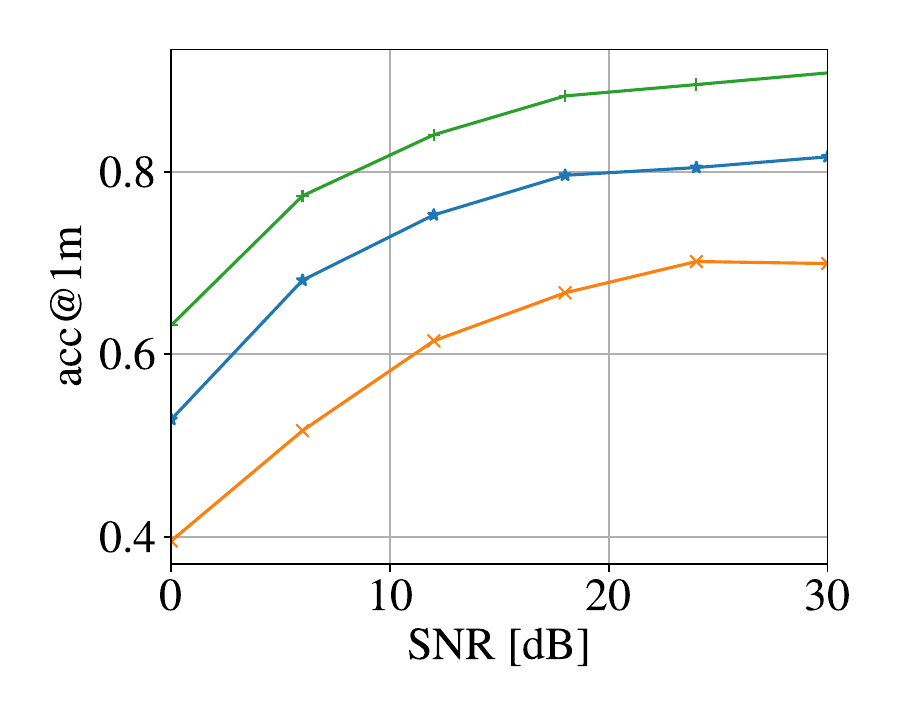}
                \caption{Accuracy vs SNR}
        \end{subfigure}\hfill
        \begin{subfigure}[b]{0.195\linewidth}
                \centering
                \includegraphics[width=\linewidth, trim={0.5cm 0.9cm 0 0.8cm},clip]{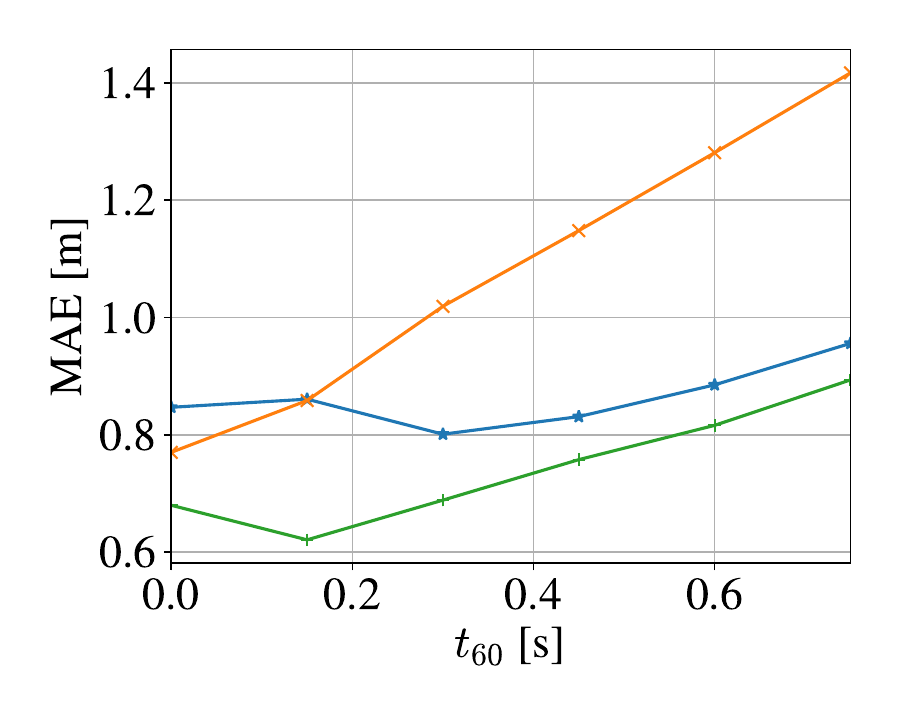}
                \caption{MAE vs $t_{60}$}
        \end{subfigure} \hfill
        \begin{subfigure}[b]{0.195\linewidth}
                \centering
                \includegraphics[width=\linewidth, trim={0.5cm 0.9cm 0 0.8cm},clip]{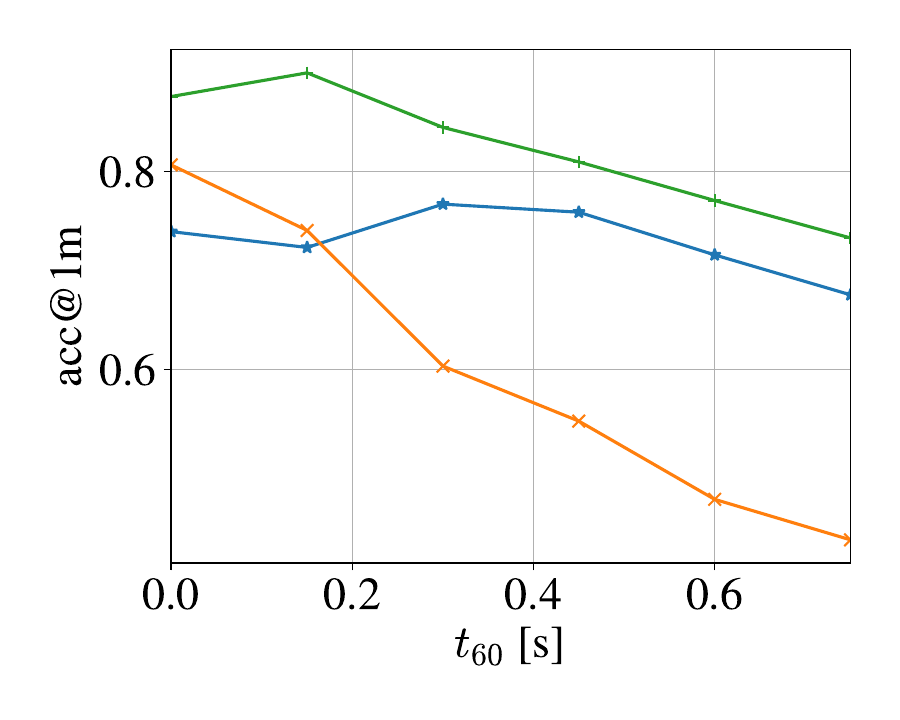}
                \caption{Accuracy vs $t_{60}$}
        \end{subfigure}
         \begin{subfigure}[b]{0.195\linewidth}
                \centering
                \includegraphics[width=\linewidth, trim={0.5cm 0.5cm 0 0.8cm},clip]{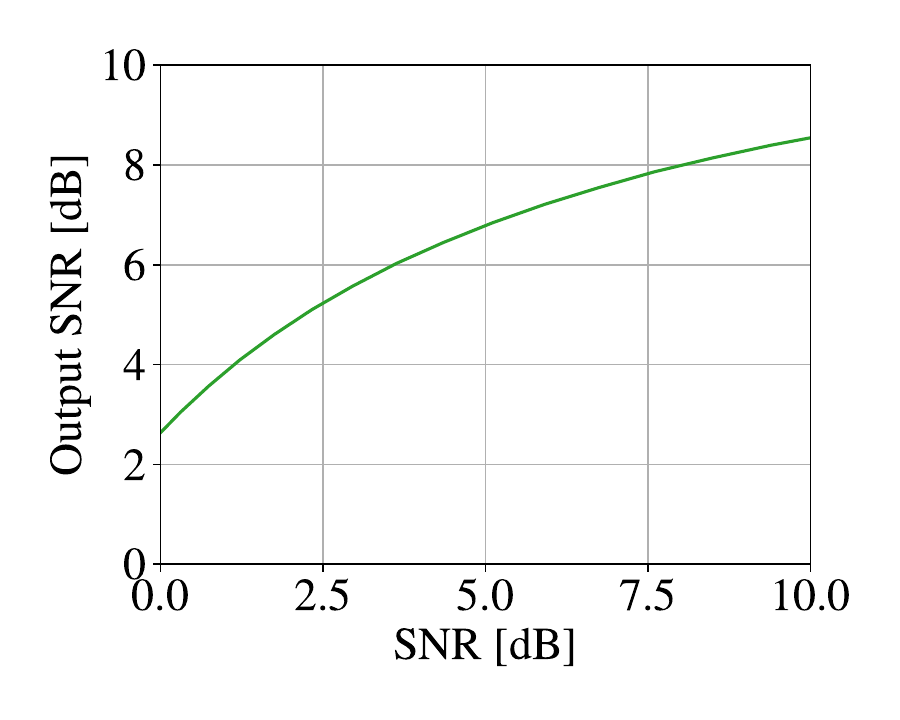}
                \caption{Output vs input SNR}
        \end{subfigure}
        \caption{Results on the simulated dataset under varying noise and reverberation conditions.}
        \label{simulated-errors}
\end{figure*}

Since DI-NN does not support a variable number of microphones, we train separate models DI-NN$_M$ for each scenario where $M$ microphones are used. The masked version of GNN is trained by randomly sampling a subset containing between 5 and 11 of microphones, and  we denote this method GNN$_\mathcal{M}$. However, unlike our proposed method, these models cannot predict microphone locations, or exploit the sound source audio.

The results using different number of known microphone coordinates are shown in Table \ref{mic_ablation}, where it can be seen that our method is consistently more robust when performing localization using a subset of the microphones. In addition, our method can also be evaluated in the scenario where the sound $\mathbf{s}_0$ emitted from the source is given (and hence audio from the 12:th microphone is not masked), but its location unknown (Setup $1_b$). Evaluating the same model in this setup shows that it can exploit the additional data to improve localization performance. Furthermore, our method can exploit audio from microphones in unknown locations, denoted as Setup $2_a$ (unknown source audio) and $2_b$ (known source audio), where audio from all microphones are used as input, but their coordinates are masked (except for the $M$ known). This improves the localization performance and also allows for localization of the microphones themselves. Table \ref{mic_loc} shows that this is possible for a small number of unknown microphone locations, but the errors become very large when less than 8 microphone locations are known.

\paragraph{Ablation study.} Next, we ablate the different components of our method in Table \ref{ablation}. Notably, the pairwise positional encoding is crucial for the method to work, since it allows the transformers to connect audio and coordinates from the same microphone. The ablation also shows that our random masking strategy significantly improves robustness to missing microphones. Providing TDOA features as an additional input drastically reduces the localization error, and we note that using a pre-trained NGCC-PHAT yields significantly better results compared to regular GCC-PHAT inputs.

\begin{table}[t]
\centering
\footnotesize
\caption{Sound source localization MAE [cm] on the \traj{speech3} trajectory using subsets of microphones and network modules.}
\begin{tabular}{l|l|cc} 
\toprule
Setup & Method & $M=7$ & $M=11$ \\ \midrule
$1_a$ & wav2pos baseline & $148.8 \pm 2.4$ & $144.3 \pm 3.1$ \\
& +pairwise pos-enc.\ & $129.7 \pm 2.9$ & $60.1 \pm 2.5$ \\
& +random masking & $96.7 \pm 3.0$ & $81.2 \pm 3.0$ \\
& +max-pooling & $92.8 \pm 2.9$ & $73.4 \pm 3.8$  \\
& +TDOA feat.\ (GCC-PHAT) & $57.9 \pm 1.8$ & $32.9 \pm 1.3$ \\
& \hphantom{+TDOA feat.} (NGCC-PHAT) & $\mathbf{38.8 \pm 1.7}$ & $\mathbf{22.7 \pm 1.1}$  \\ \bottomrule %
\end{tabular}
\label{ablation}
\end{table}

\paragraph{Simulated environment.} So far, we have only considered recordings with microphones in a limited number of possible locations. In order to validate our approach in scenarios with flexible microphone locations and changes in signal to noise ratio (SNR) and reverberation times, we perform additional experiments in a controlled simulated environment\footnote{Code: \url{https://github.com/axeber01/wav2pos/}}. We use the same setup $1_a$ as for LuViRA, but microphone locations are randomly sampled across each of the walls, floor and ceiling, for a total of six microphones. We use recordings from the LibriSpeech dataset \cite{panayotov2015librispeech} and create a 20\thinspace000/2\thinspace000 train/test split based on speaker-ids. Results are shown in Figures \ref{simulated-errors}a-d, where performance is evaluated over a range of SNRs and reverberation times. Notably, our proposed method outperforms DI-NN and GNN in all scenarios. GNN, which relies on GCC-PHAT as input, performs poorly as reverberation increases, which highlights the necessity of using better feature extractors for good performance in such conditions. In addition, Figure \ref{simulated-errors}e shows the signal de-noising performance of our method, yielding positive gains roughly in the range of 0 to 7 dB. However, de-noising performance is limited by the embedding dimension of the encoder and decoder, making it difficult to reconstruct high-frequency content.

\section{Conclusions and Future Work}
In this work, we have proposed a general SSL method that can be used in a wide range of problem scenarios. We conjecture that our method can also be further extended to localizing multiple sound sources. This is possible due to the flexibility of masked autoencoders, where additional inputs or outputs can be added seamlessly. It is also possible to consider the full self-calibration problem where no microphone locations are known, but this requires processing longer sequences of moving sound sources. We leave this as future work and hope that it can inspire the wider research community to create more challenging localization datasets and tasks.

\bibliographystyle{IEEEtran}
\bibliography{wav2pos}

\appendix
\paragraph{Details on model training.} The implementations of DI-NN\footnote{\url{https://github.com/egrinstein/di_nn/}} and GNN\footnote{\url{https://github.com/egrinstein/gnn_ssl/}} available online are extended from two to three dimensions. We consider using GNN with spatial likelihood functions infeasible in large 3D environments, thus we use GNN with GCC-PHAT inputs only. We also tried pre-trained NGCC-PHAT features as input to GNN, but the training then failed to converge. 

For $M$ microphones GNN uses $M(M - 1)/2$ TDOA features by only computing $\mathbf{R}_{ij}$ and not $\mathbf{R}_{ji}$, which speeds up computation time, but breaks permutation invariance. In contrast, wav2pos uses all $M(M-1)$ features in order to preserve this property. This also results in slightly better localization performance. Computation time is saved by noting that $\mathbf{R}_{ij}[t] = \mathbf{R}_{ji}[-t]$ for any time delay $t$, a property that holds for both GCC-PHAT and NGCC-PHAT.

NGCC-PHAT is trained using an available implementation\footnote{\url{https://github.com/axeber01/ngcc/}} with slight modifications. At pre-training time, two random microphones are picked for each training example and the TDOA is estimated using classification in the range of integers $-\tau, ..., \tau$ using the cross-entropy loss. The maximum possible time delay between two microphones can be calculated by considering the distance between the two most separated microphones. For the LuViRA dataset, this results in a maximum delay of $\tau = 314$ samples. At inference time, TDOA estimates are computed for all pairs of microphones.

The multilateration method runs in Matlab using open source code\footnote{\url{https://github.com/kalleastrom/StructureFromSound2/}}. We modify the code to not consider the full self-calibration problem, but only the sound source localization with known microphone positions. This method uses a RANSAC loop that tests each of the four strongest peaks in the GCC-PHAT feature per microphone pair.. When adopting the method to use NGCC-PHAT, we only consider a single peak, since NGCC-PHAT is trained to estimate a single TDOA.

Model training is done in Pytorch on a single NVIDIA A100 GPU. Training and inference times are shown in \autoref{inference}. We do not report time for multilateration, as there is no training step and the inference code was not optimized for speed. The wav2pos network without TDOA features corresponds to the \autoref{ablation} max-pooling step. We note that computing TDOA features using NGCC-PHAT increases the execution time of our method considerably, since these are computed sequentially for all 55 pairwise microphone combinations. This could be parallelized in order to speed up execution time.

\begin{figure}[ht]
\centering
\includegraphics[width=0.9\linewidth, trim={0.5cm 1.0cm 0 2cm},clip]{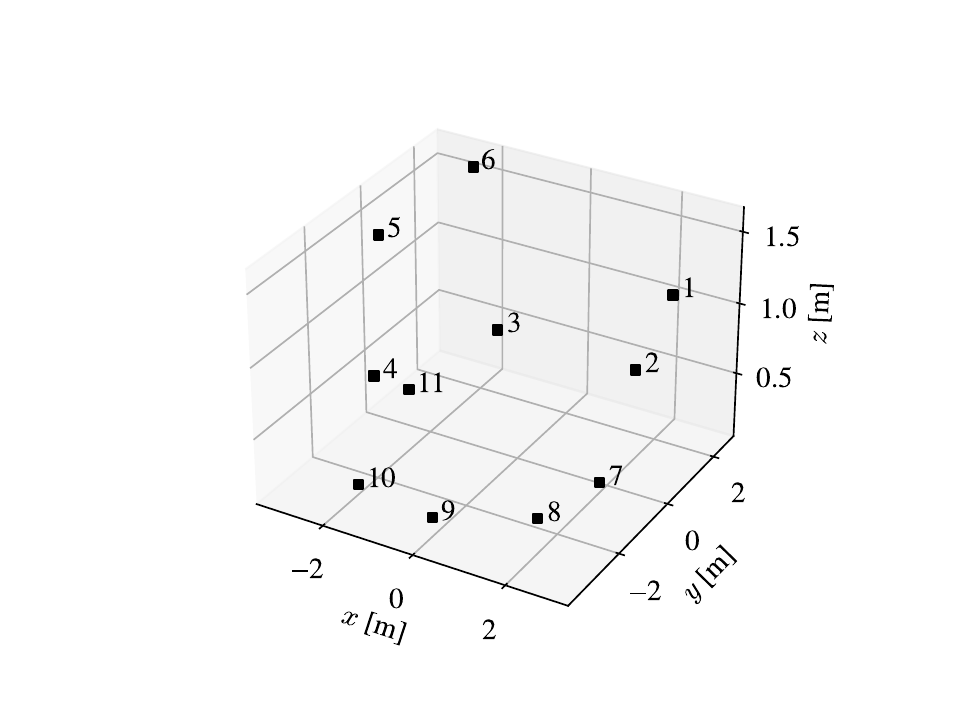}
\caption{Microphone locations in the LuViRA dataset.}
\label{luvira-room}

\includegraphics[width=0.9\linewidth, trim={0.5cm 1.0cm 0 2cm},clip]{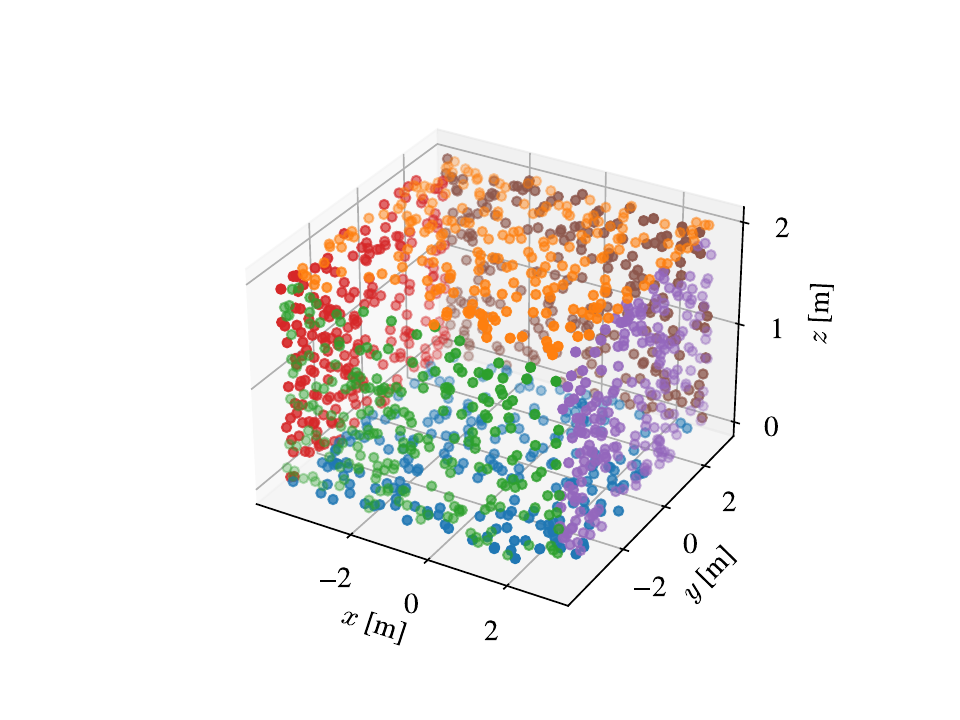}
\caption{Microphone distribution in the simulated dataset. Each dot shows a microphone for a single training example. The color shows which of the six faces of the room it belongs to.}
\label{sim-room}
\end{figure}

\paragraph{Dataset details.} Figure \ref{luvira-room} shows the 3D layout of the microphones in the LuViRA dataset, which are identical for all eight recordings. In addition, the microphone locations in simulated dataset are visualized in Figure \ref{sim-room}. Table \ref{sim-data} provides the train/test split used for the simulated dataset.

\begin{figure*}
\centering
        \begin{subfigure}[t]{0.24\linewidth}
                \centering
                \includegraphics[width=\linewidth, trim={0.5cm 0.5cm 0 0},clip]{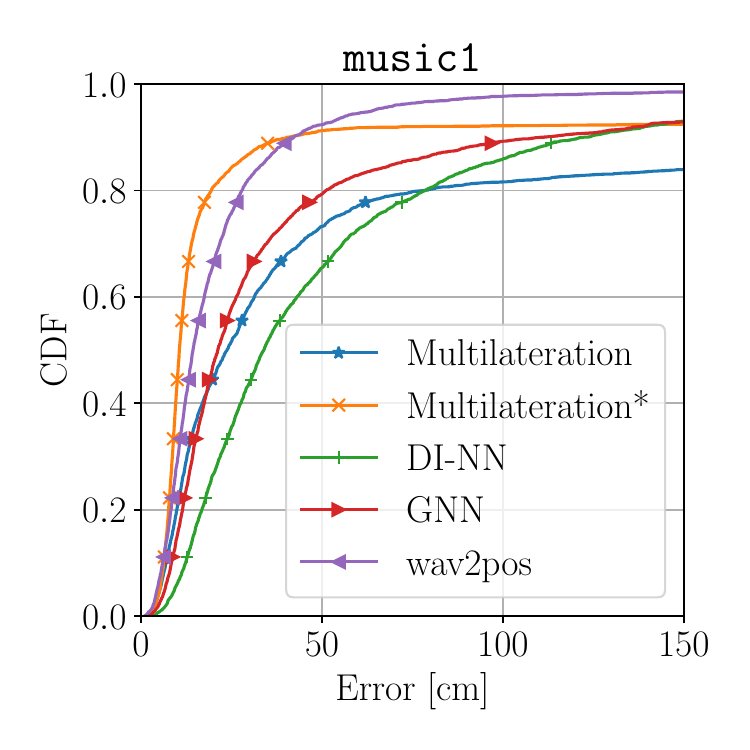}
        \end{subfigure}\hfill
        \begin{subfigure}[t]{0.24\linewidth}
                \centering
                \includegraphics[width=\linewidth, trim={0.5cm 0.5cm 0 0},clip]{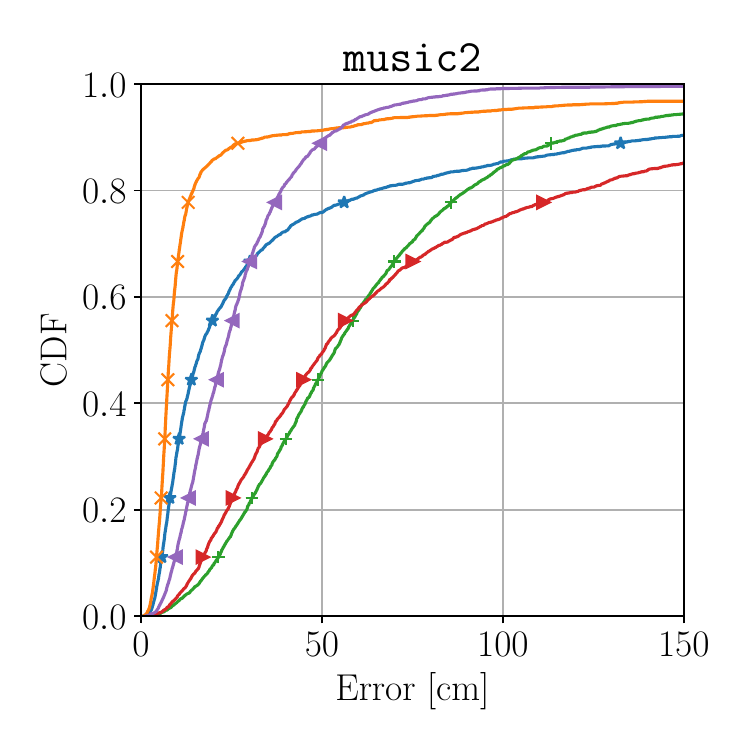}
        \end{subfigure}\hfill
                \begin{subfigure}[t]{0.24\linewidth}
                \centering
                \includegraphics[width=\linewidth, trim={0.5cm 0.5cm 0 0},clip]{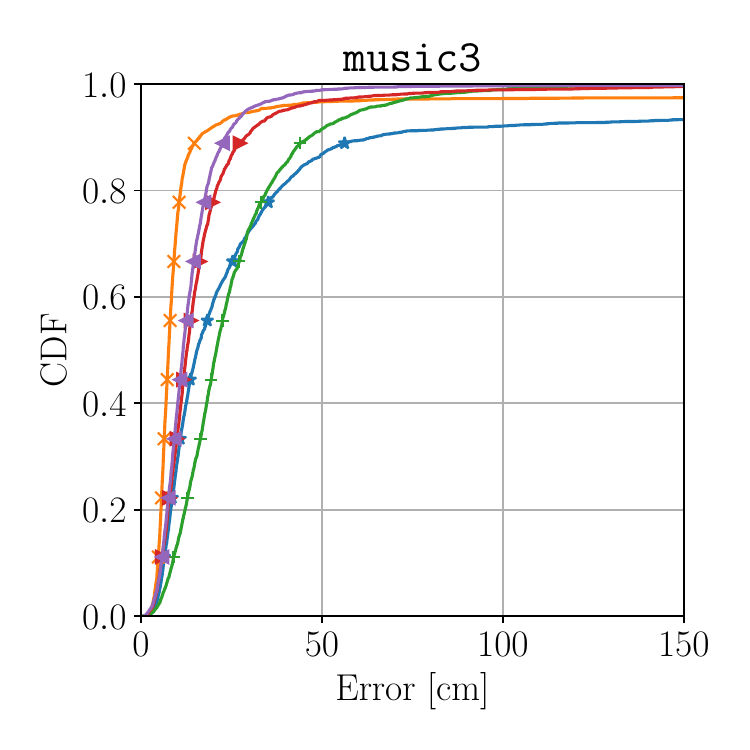}
        \end{subfigure}\hfill
        \begin{subfigure}[t]{0.24\linewidth}
                \centering
                \includegraphics[width=\linewidth, trim={0.5cm 0.5cm 0 0},clip]{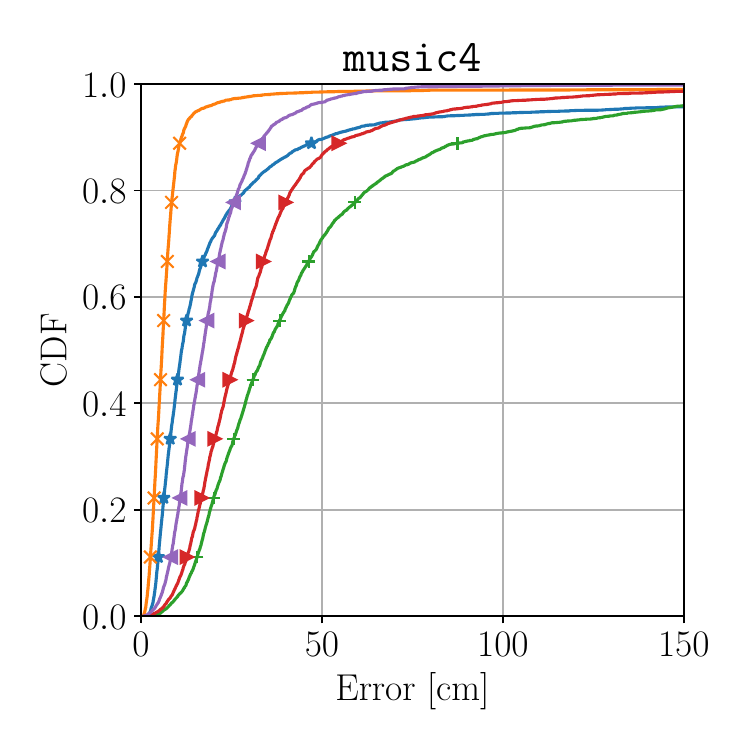}
        \end{subfigure}\hfill
        \begin{subfigure}[t]{0.24\linewidth}
                \centering
                \includegraphics[width=\linewidth, trim={0.5cm 0.5cm 0 0},clip]{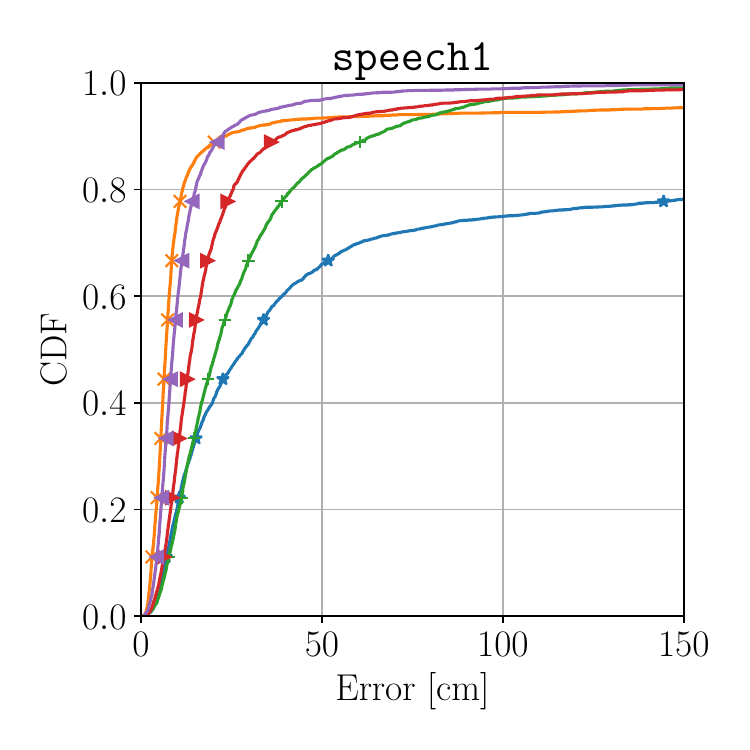}
        \end{subfigure}\hfill
        \begin{subfigure}[t]{0.24\linewidth}
                \centering
                \includegraphics[width=\linewidth, trim={0.5cm 0.5cm 0 0},clip]{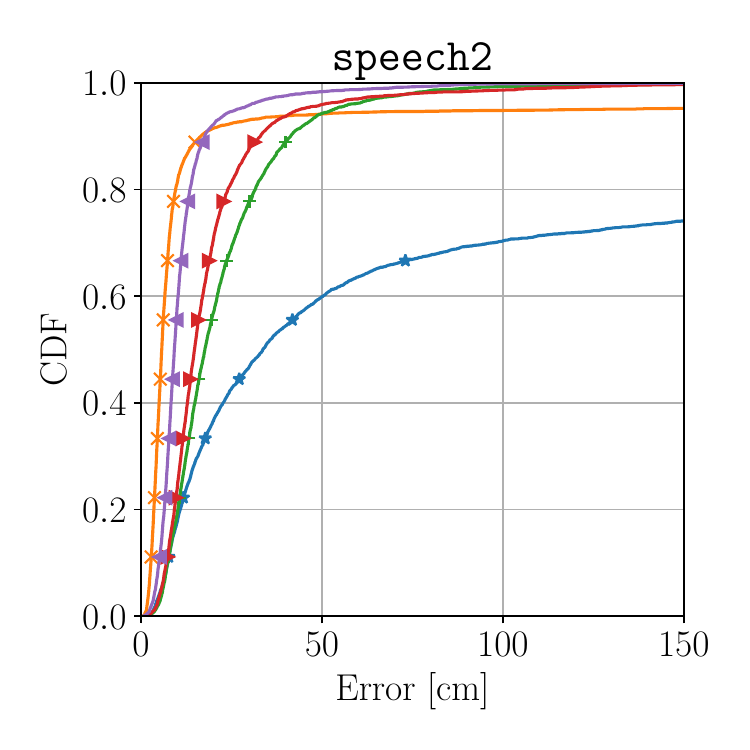}
        \end{subfigure}\hfill
        \begin{subfigure}[t]{0.24\linewidth}
                \centering
                \includegraphics[width=\linewidth, trim={0.5cm 0.5cm 0 0},clip]{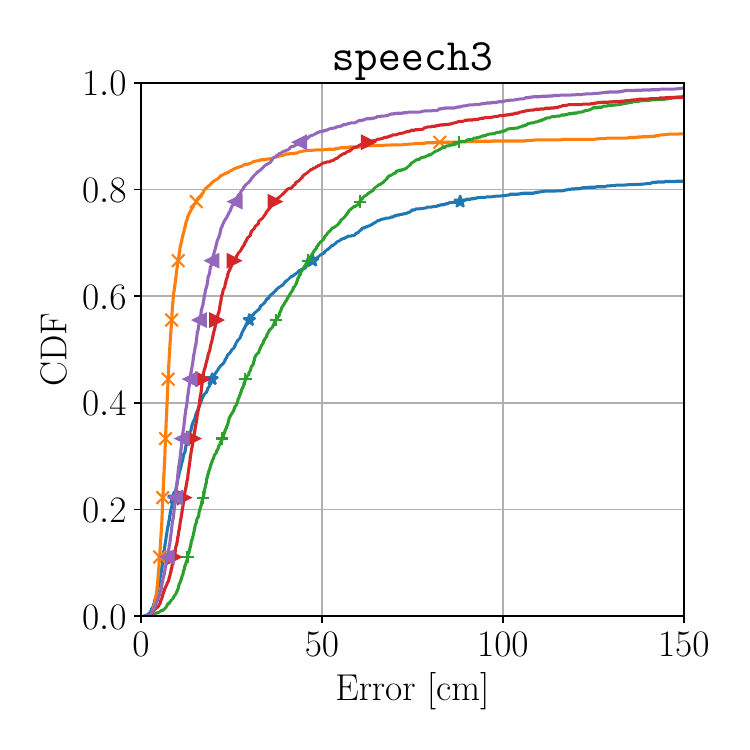}
        \end{subfigure}\hfill
        \begin{subfigure}[t]{0.24\linewidth}
                \centering
                \includegraphics[width=\linewidth, trim={0.5cm 0.5cm 0 0},clip]{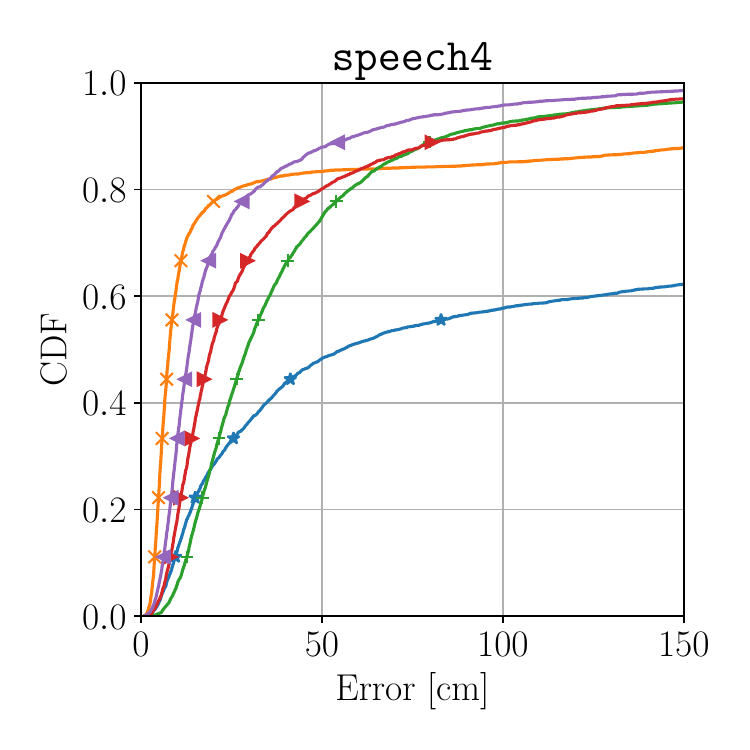}
        \end{subfigure}\hfill
        \caption{Cumulative error distributions on the LuViRA dataset using setup $1_a$ and different test splits.}
        \label{luvira-errors}
\end{figure*}

\begin{figure*}[t]
\centering
        \begin{subfigure}[t]{0.33\linewidth}
                \centering
                \includegraphics[width=\linewidth, trim={0.5cm 0 0 0},clip]{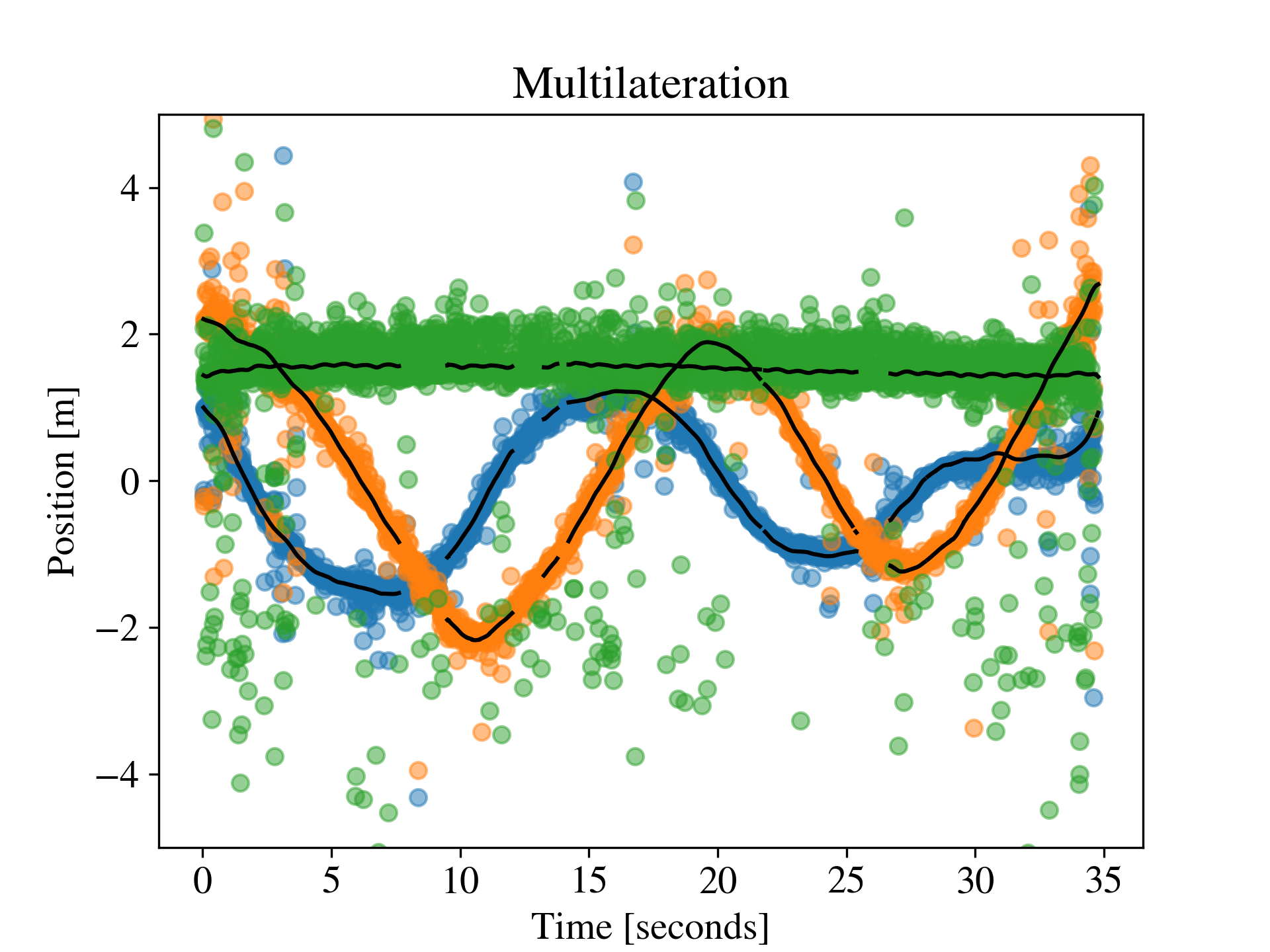}
        \end{subfigure}\hfill
        \begin{subfigure}[t]{0.33\linewidth}
                \centering
                \includegraphics[width=\linewidth, trim={0.5cm 0 0 0},clip]{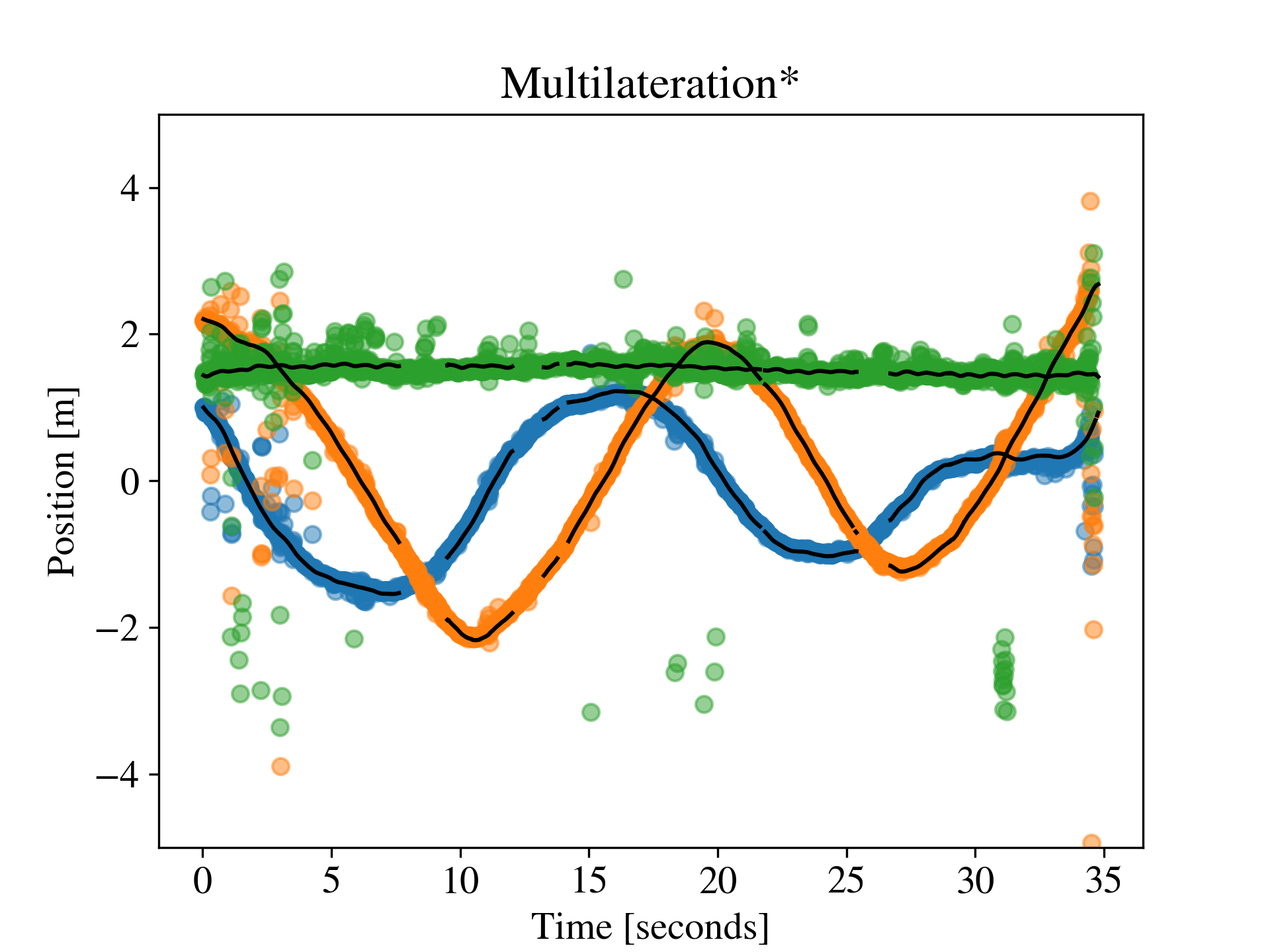}
        \end{subfigure}\hfill
        \begin{subfigure}[t]{0.33\linewidth}
                \centering
                \includegraphics[width=\linewidth, trim={0.5cm 0 0 0},clip]{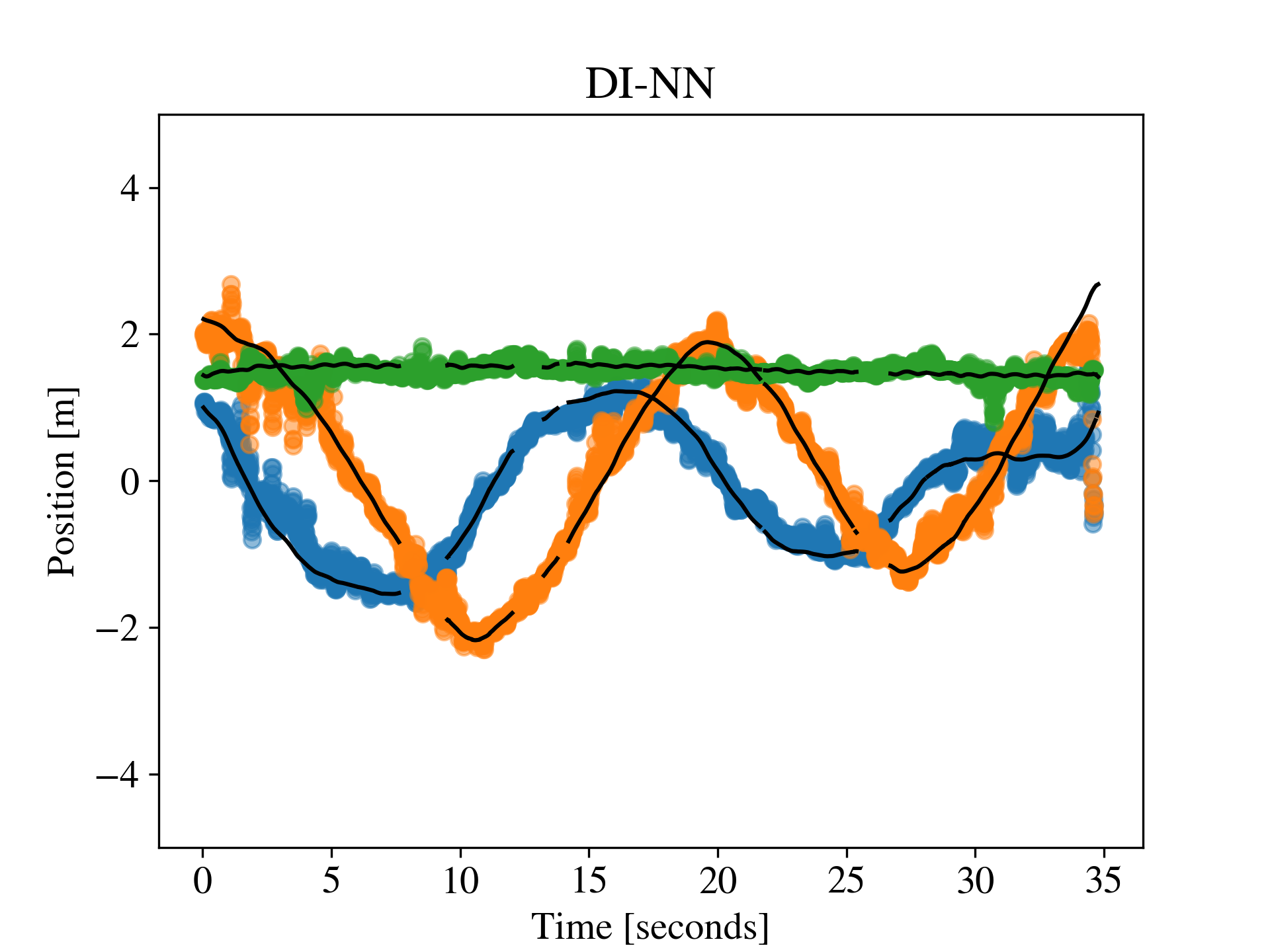}
        \end{subfigure} \hfill
        \begin{subfigure}[t]{0.33\linewidth}
                \centering
                \includegraphics[width=\linewidth, trim={0.5cm 0 0 0},clip]{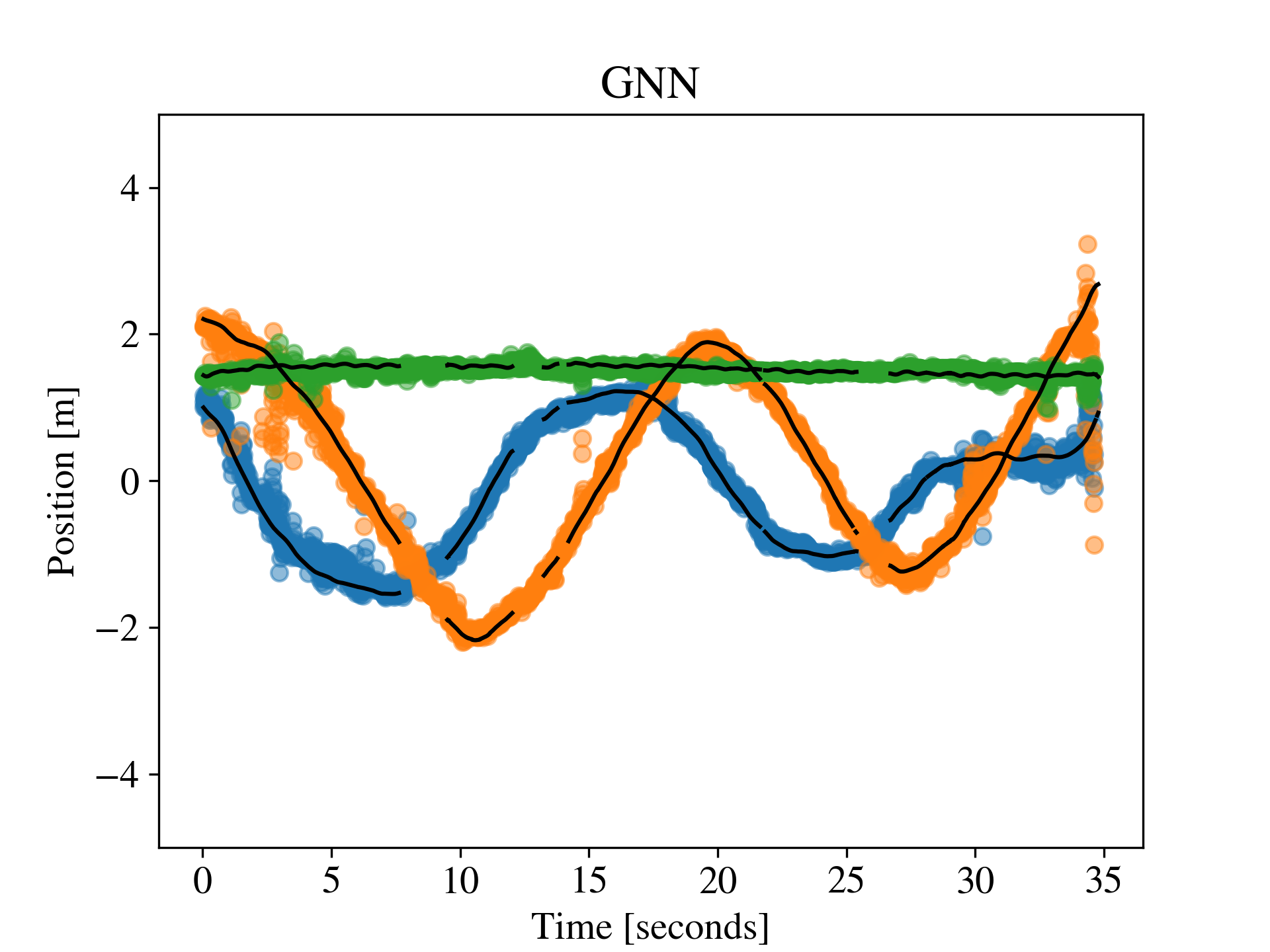}
        \end{subfigure}
         \begin{subfigure}[t]{0.33\linewidth}
                \centering
                \includegraphics[width=\linewidth, trim={0.5cm 0 0 0},clip]{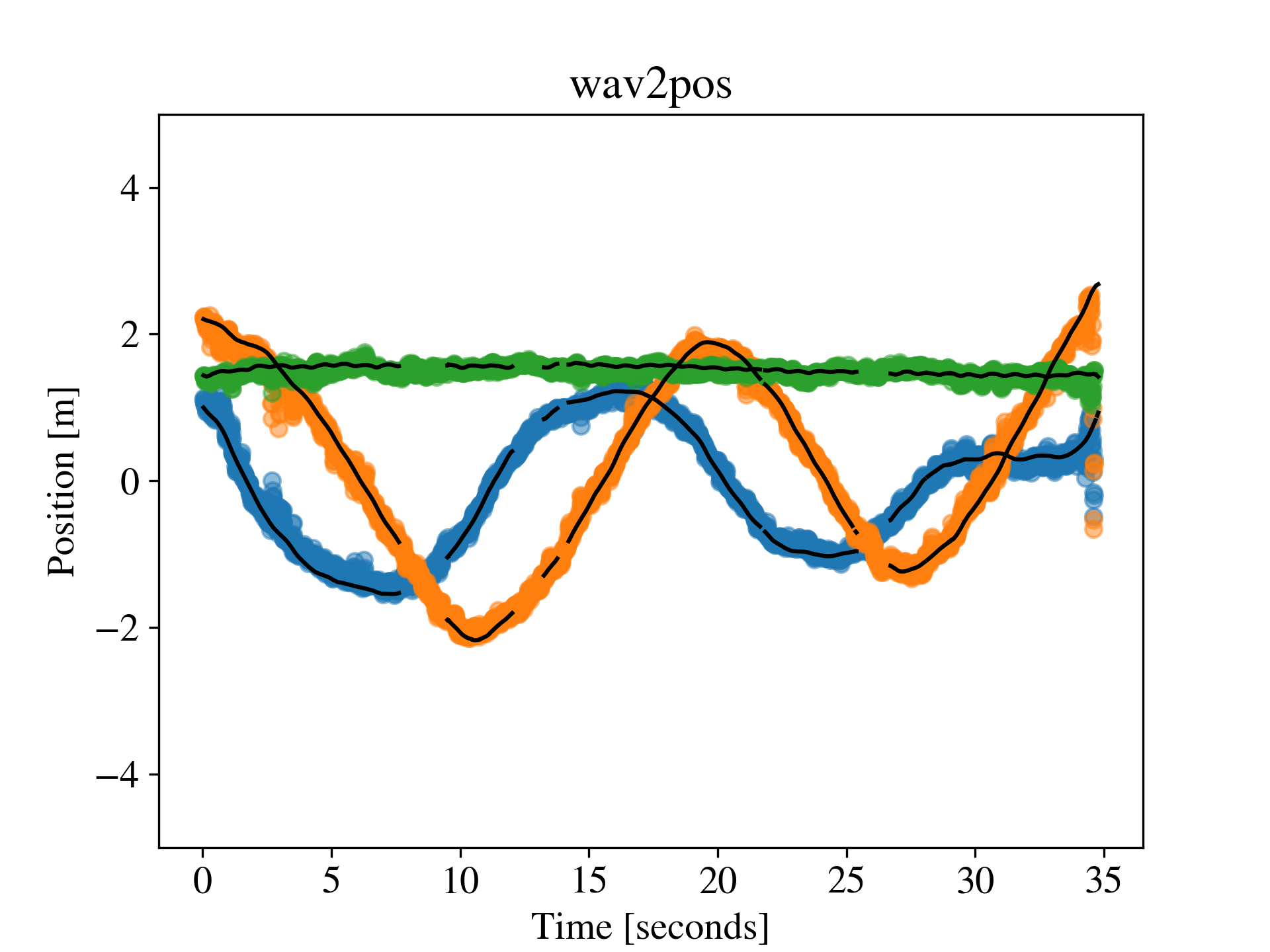}
        \end{subfigure}
        \caption{Visualizations of the predictions on the \traj{music3} track, where each coordinate prediction is shown separately ($x$: blue, $y$: orange, $z$: green). Ground truth coordinates are traced in black (some timestamps are missing ground truth).}
        \label{xyz-pred-music}
\end{figure*}

\paragraph{Additional results on LuViRA dataset.} We provide additional results for all methods on several splits of the LuViRA
dataset in \autoref{all-splits}. The train/test splits are constructed such that six tracks are used for training, e.g.\ \traj{music1-3} and \traj{speech1-3}, and two are used for testing, e.g.\ \traj{music4} and \traj{speech4}. Out of the eight tracks, \traj{music2} and \traj{music4} are the only ones with significant height variation in the source trajectory. The corresponding cumulative error distributions are shown in Figure \ref{luvira-errors} and additional visualizations of model predictions are shown in Figure \ref{xyz-pred-music}. It can be seen that although the multilateration methods are often very accurate, they have a long tail in the error distributions due to outliers, whereas the learning-based methods do not suffer from this problem. 

\begin{table}[t!]
\centering
\footnotesize
\caption{Training and inference times for setup $1_a$ with $M=11$ microphones, measured on a single A100 GPU. Training was done for 500 epochs on the LuViRA dataset.}
\begin{tabular}{l|c|c} \toprule
Method & Training [h] & Inference  [ms] \\ \midrule
DI-NN & 29.5 & 0.2 \\ 
GNN & 34.7 & 1.0 \\
wav2pos w/o TDOA feat. & 27.8 & 0.2 \\
wav2pos & 64.3 &  5.9 \\ \bottomrule
\end{tabular}
\label{inference}
\vspace{1em}
\centering
\footnotesize
\caption{Speaker-ids used in the simulated dataset. Recordings shorter than one second are discarded.}
\begin{tabular}{c|c|c} \toprule
Dataset & Test speaker-ids & Train speaker-ids\\ \midrule
Librispeech test-clean & 61, 121, 237 & all other 43 speakers \\ \bottomrule
\end{tabular}
\label{sim-data}
\end{table}

\begin{table*}[t]
\centering
\footnotesize
\caption{Mean absolute error [cm] on the LuViRA dataset using setup $1_a$ and different test splits.}
\begin{tabular}{l|c|c|c|c|c|c|c|c}
\toprule
Method & \traj{music1} & \traj{music2} & \traj{music3} & \traj{music4} & \traj{speech1} & \traj{speech2} & \traj{speech3} & \traj{speech4} \\ \midrule
Multilat & $67.2 \pm 3.2$ & $48.6 \pm 2.3$ & $38.8 \pm 2.5$ & $28.1 \pm 1.6$ & $80.9 \pm 3.2$ & $90.4 \pm 3.5$ & $72.8 \pm 4.4$ & $124 \pm 3.8$  \\
Multilat* & $32.9 \pm 2.3$ & $\mathbf{20.7 \pm 1.5}$ & $16.3 \pm 1.6$ & $\mathbf{9.5 \pm 0.8}$ & $20.7 \pm 1.7$ & $18.9 \pm 1.6$ & $34.9 \pm 3.2$ & $41.9 \pm 2.5$ \\
DI-NN & $54.0 \pm 1.9$ & $64.9 \pm 1.5$ & $26.0 \pm 0.8$ & $47.1 \pm 1.2$ & $29.6 \pm 0.9$ & $22.5 \pm 0.6$ & $44.7 \pm 1.7$ & $43.2 \pm 1.5$ \\
GNN & $42.3 \pm 1.8$ & $74.9 \pm 2.1$ & $17.0 \pm 0.7$ & $33.6 \pm 0.8$ & $21.6 \pm 0.9$ & $19.2 \pm 0.6$ & $31.9 \pm 1.6$ & $35.1 \pm 1.3$ \\
wav2pos & $\mathbf{22.7 \pm 1.0}$  & $28.1 \pm 0.8$ & $\mathbf{14.2 \pm 0.5}$ & $19.9 \pm 0.4$ & $\mathbf{13.1 \pm 0.6}$ & $\mathbf{11.7 \pm 0.4}$ & $\mathbf{23.6 \pm 1.2}$ & $\mathbf{24.8 \pm 0.9}$ \\ \bottomrule
\end{tabular}
\label{all-splits}
\end{table*}

\end{document}